\documentclass[twocolumn,aps,prb,showpacs,preprintnumbers]{revtex4}%
\usepackage{mathrsfs}
\usepackage{amsmath}
\usepackage{amsfonts}
\usepackage{amssymb}
\usepackage{amsthm}
\usepackage{graphicx}
\usepackage{color}
\usepackage{graphics}
\usepackage{epsfig}
\providecommand{\U}[1]{\protect\rule{.1in}{.1in}}
\usepackage{amsmath}
\usepackage{ulem}
\usepackage{amsfonts}
\usepackage{amssymb}
\usepackage{amsthm}
\usepackage{graphicx}
\usepackage{epsfig}
\usepackage{color}
\usepackage{subfigure}
\usepackage{graphics}

\begin{document}
\title{Anti-Levitation in the Integer Quantum Hall Systems}

\author{C. Wang$^{1,2}$}
\author{Y.  Avishai$^{1,3}$}
\email[corresponding author: ]{yshai@bgumail.bgu.ac.il}
\author{Yigal Meir$^{3}$}
\author{X. R. Wang$^{1,2}$}
\email[corresponding author: ]{phxwan@ust.hk}
\affiliation{$^{1}$Physics Department, The Hong Kong University of
Science and Technology, Clear Water Bay, Kowloon, Hong Kong}
\affiliation{$^{2}$HKUST Shenzhen Research Institute, Shenzhen 518057,
China}
\affiliation{$^{3}$Department of Physics, Ben-Gurion University of
the Negev Beer-Sheva, Israel}
\date{\today}
\begin{abstract}
The evolution of extended states of two-dimensional electron gas
with white noise randomness and field is numerically investigated
by using the Anderson model on square lattices. Focusing on the
lowest Landau band we establish an {\it anti-levitation scenario}
of the extended states: As either the disorder strength $W$ increases
or the magnetic field strength $B$ decreases, the energies of the
extended states move {\it below} the Landau energies pertaining to
a clean system. Moreover, for strong enough disorder, there is a
disorder dependent critical magnetic field $B_c(W)$ below which
there are no extended states at all. A general phase diagram in the
$W-1/B$ plane is suggested with a line separating domains of localized
and delocalized states.
\end{abstract}
\pacs{71.30.+h, 73.20.Jc}
\maketitle
\section{Introduction}
\label{I}
Energies of an electron in a {\it clean} two-dimensional system subject to a
strong perpendicular magnetic field $B$ form sharp Landau levels at energies
$\varepsilon_n=(n+1/2)\hbar \omega_c$, $n=0,1,2,\ldots$ [where $\omega_c=eB/
(m c)$] and the corresponding eigenstates (Landau functions) are extended.
If the system is moderately disordered, for example by a white-noise random
on-site energy of zero mean and fluctuation strength $W$ such that
$W<\hbar \omega_c$, the Landau levels are broadened to form separated
Landau bands (LBs).\cite{Huckestein} The density of states of each LB is
maximal around its center $E_n$.\cite{Wegner} While the (possibly
degenerate) eigenstates at energy $E_n$ are still extended, states at
energies $\varepsilon\ne E_n$ are localized. This is the origin of the
integer quantum Hall effect (IQHE).\cite{QHE} The question of whether
$E_n=\varepsilon_n$ is one of the topics discussed in this work. Each LB
$i$ is characterized by a topological (Chern) integer $\nu_i$,\cite{TNN}
and the Hall conductivity $\sigma_{xy}$ at Fermi energy
$E_n<\varepsilon_F <  E_{n+1}$ is equal to $\sum_{i=1}^n \nu_i$ in the unit
of $e^{2}/h$. Strictly speaking, only extended states at energy $E_i$
within each LB contribute to its Chern number. Chern numbers cannot be
created or destroyed by an adiabatic change of $B$ or $W$.

One of the fundamental issues in the physics of the IQHE is to elucidate
the evolution of extended states in LBs with stronger disorder and/or
weaker magnetic field such that the inequality $W<\hbar \omega_c$ is no
longer strictly satisfied. As $B \to 0$ all electronic states in a disordered
two-dimensional system are localized \cite{Abrahams} and for $B=0$
there are neither LB nor Chern numbers. On the other hand, a LB with Chern
number $\nu$ cannot lose it unless it is annihilated by an opposite Chern
number $-\nu$ belonging to another LB. In Refs.~\onlinecite{KLZ,QNiu1,GXiong}
the scenario of Chern number annihilation as $B \to 0$ has been discussed
on a qualitative level.
In order to add more quantitative perception, it is vital to elucidate the
behavior of extended states in LBs as the magnetic field gradually decreases
to zero, or as the disorder gradually increases. Different answers to this
question lead to different global phase diagrams \cite{KLZ,QNiu1,GXiong} for
the IQHE.


In the absence of spin-orbit interaction, the prevailing paradigm
(assuming 2D continuum geometry) is that when $B \to 0$, all extended
states float up to infinite energy. The quantitative form of this
{\it levitation scenario} \cite{Laughlin,Khmelnitskii} states that the
energy of an extended state in the $n$th LB goes like
\begin{align}
\begin{split}
E_{n}=\varepsilon_n[1+(\omega_{c}\tau)^{-2}],
\end{split}
\label{levitation}
\end{align}
where $\tau$ is the impurity scattering time. Thus, extended states float
upward as $\omega_c\rightarrow 0$ or $\tau\rightarrow 0$.\cite{Haldane}
As far as experiments are concerned, the levitation scenario is still not
settled. Some experiment support it \cite{HWJiang,TWang,Glozman,Kravechenko}
and others do not.\cite{Shahar,Shahar1} Some theoretical works based on
continuous 2D geometry treat the levitation scenario using numerous
approximation methods as well as various numerical calculations.
\cite{Haldane,Shahbazyan,  Kagalovsky1, Kagalovsky2, Fogler} Most of them
support the general idea although there is no strict evidence that extended
states float up to infinity. Numerical simulations on a lattice are also not
conclusive. For white-noise random on-site energy, levitation is not
substantiated \cite{QNiu1} while for finite range correlated disorder, weak
levitation is predicted.\cite{HSong,DNSheng,Pereira}

\begin{figure}
  \begin{center}
  \includegraphics[width=8cm]{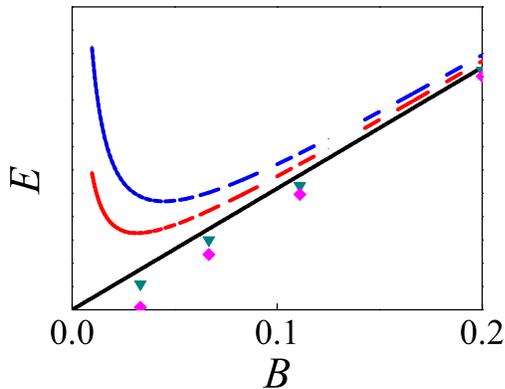}
  \end{center}
\caption{(Color online) Schematic picture of the anti-levitation scenario
(obtained from a solution of the lattice model with white-noise
distribution of site disorder), contrasted with the levitation scenario
formulated in Eq.~(\ref{levitation}). Energies of extended states on the
lowest LB are shown as a function of the magnetic field for different cases.
(1) Clean system in the continuous geometry (solid line) where the energy
follows the relation $E_0=\frac{1}{2}\hbar \omega_c$.
(2)  Disordered systems in the continuous geometry following Eq. \ref
{levitation} for $\tau_1$ (lower dashed line) $> \tau_2$ (upper dashed line).
(3) Disordered system in the lattice geometry with white-noise random
on-site energy for disorder strengths $W_1$ (triangles) $< W_2$ (diamonds).
Anti-levitation occurs for fixed $B$ with increasing $W$ and for fixed $W$
with decreasing $B$.}
\label{PD}
\end{figure}

In this paper we revisit this issue by focusing on the evolution of
extended states in the lowest LB ($n=0$) using a square lattice geometry and
white-noise random on-site energy. The lattice geometry is especially useful at
low magnetic field where LBs strongly overlap. Our main result
(to be substantiated below) is that
under these conditions, extended states in the lowest LB
{\it plunge  down instead of floating up} when $\omega_c \tau \to 0$. This anti-levitation
behavior is schematically  displayed in Fig.~\ref{PD} and contrasted with
the levitation scenario encoded in Eq.~(\ref{levitation}).
This
{\it anti-levitation scenario} is consistent with the concept of level
repulsion as explained below. It is substantiated by two independent numerical
approaches. In the first one, the extended state energy is identified as the
critical energy for the IQHE plateaux transition where the localization length
diverges. The second one is based on the calculation of the
participation ratio (PR). \\

This paper is organized as follows. In the first part of Sec.\ref{II},
the model describing an electron on a 2D square lattice with white-noise
disorder under a perpendicular magnetic field is very briefly discussed.
In the second part of Sec.\ref{II} we give, for the sake of
self-consistence, a short explanation of the transfer matrix and PR
methods designed to locate the extended state energies $E_n$.
Section \ref{III} is devoted to presentation of the numerical results
and their analysis, while a short summary is presented in Sec.\ref{IV}.

\section{Model and Methods}
\label{II}
We consider a tight-binding Hamiltonian on a square lattice,
\begin{align}
\begin{split}
H=\sum_{i}\epsilon_{i}a_{i}^{\dagger}a_{i}+\sum_{<ij>}
\exp(i\phi_{ij})a_{i}^{\dagger}a_{j}.
\end{split}
\label{Hamiltonian}
\end{align}
Here $i=(n_ia,m_ia)$ is a point on a square lattice of lattice constant $a$, where
$1 \le n_i \le L$ and $1 \le m_i \le M$ ($n_i,m_i,L,M$ are nonnegative integers),
and $a_i^\dagger, a_i$ are electron creation and annihilation operators on site $i$.
The on-site energy $\epsilon_{i}$ on site $i$ is a random number uniformly
distributed in the range of $[-\frac{W}{2},\frac{W}{2}]$. Thus, $W$ measures the
degree of randomness.
The symbol $<ij>$ indicates that $i$ and $j$ are nearest-neighbor sites.
The magnitude of the hopping coefficient (prefactor of the exponent) is used as
an energy unit. The magnetic field is introduced through the Peierls substitution
\cite{xrw1} by adding a phase $\phi_{ij}=2\pi(e/h)\int_{i}^{j}\vec{A}\cdot d\vec{l}$
to the hopping coefficient. The vector potential is chosen as $\vec{A}=(0,Bx,0)$ for
a uniform magnetic field $\vec{B}$ along the $z-$direction, where $B$ is in
the units of flux quantum $e/h$ per plaquette. Under this gauge, the nonzero phase
exists only on bonds along the $y-$direction.
The energy range of eigenstates for a pure system is $[-4,4]$. In the presence of
mild disorder potential, the energy spectrum slightly extends beyond $[-4,4]$.
Because the model \eqref{Hamiltonian} has particle-hole symmetry,
the discussion below can be restricted within the energy range $[-4,0]$.

To locate the extended state energies at a given magnetic field $B$ and
disorder fluctuation energy range $W$, we use two independent approaches.
The first one is the transfer matrix method. The electron is scattered from
a quasi-one-dimensional lattice (a strip) of length $L a \to \infty$ and
width $M a$ with $L \gg M$. Assume that the $x-$axis lies along the
longitudinal direction of the strip and the $y-$axis along its transverse
direction; the transfer matrix transforms the
amplitudes of the wave function on sites $(a,m a)$ to its amplitudes
on sites $(La,ma)$ ($m=1,2,\ldots,M$). From the eigenvalues of the transfer
matrix one can effectively compute the localization length $\lambda_{M}(E)$
of the scattering state at energy $E$.

In order to avoid the edge effect,  periodical boundary conditions are imposed
on the $y-$direction. Let us denote by $\psi_n$ the vector of $M$ wave function
amplitudes on the $n^{th}$ column of the lattice, namely,
$\psi_n=(\psi_{n,1},\psi_{n,2},\ldots,\psi_{n,M})^T$. Following the method
specified in Ref.~\onlinecite{XCXie}, the $2M$ vector of wave function
amplitudes $(\psi_{n+1},\psi_n)^T$ is related to the vector
$(\psi_{n},\psi_{n-1})^T$ according to the relation
\begin{align}
\begin{split}
\begin{bmatrix}
\psi_{n+1}\\
\psi_{n}
\end{bmatrix}
=\begin{bmatrix}
E-H_{n} & -I\\
I & 0
\end{bmatrix}
\begin{bmatrix}
\psi_{n}\\
\psi_{n-1}
\end{bmatrix} \equiv T_n \begin{bmatrix}
\psi_{n}\\
\psi_{n-1}~.
\end{bmatrix}
\end{split}
\label{transfer matrix}
\end{align}
where $I$ is $M\times M$ identity matrix and the matrix $H_{n}$ is the part of
Hamiltonian related to the $n$th column. The $2M \times 2M$ transfer matrix is
given by the product $T \equiv \prod_{n=1}^L T_n$. For $L \to \infty$ the $2M$
eigenvalues of $T$ (Lyapunov exponents) can be approximated as $e^{\pm L/\xi_m(E)}$
where Re$[\xi_m(E)] \ge 0$.\cite{XCXie} The localization length $\lambda_{M}(E)$
is given by $\lambda_{M}(E)$=Max$_{m}$Re$[\xi_m(E)]$. In our calculations the
strip length is chosen to be $10^{6}$, much larger than $\lambda_M$, to take
the advantage of self-averaging.

In order to obtain the localization length $\xi(E) = \lambda_{M \to \infty}(E)$
of an infinite 2D system from the localization length $\lambda_M$ of finite
systems, we employ the single parameter scaling ansatz \cite{Mackinnon,Landau1}
implying that for a large enough system $\lambda_{M}/M$ depends only on a single
parameter $M/\xi(E)$; i.e.,
\begin{align}
\begin{split}
\frac{\lambda_{M}(E)}{M}=f\left (\frac{M}{\xi(E)} \right ).
\end{split}
\label{finitesizescaling}
\end{align}
If there is a mobility edge $E_c$ that separates localized states from
extended states, then scaling theory says
\begin{align}
\begin{split}
\xi(E)\propto |E-E_{c}|^{-\nu},
\end{split}
\label{criticalexponent}
\end{align}
where $\nu$ is a universal critical exponent depending only on
dimensionality and symmetries. According to Eq. \eqref{finitesizescaling},
$\lambda_{M}(E)/M$ of different $M$ shall all cross at $E=E_c$ when extended
states for a band or merge there if the state of energy $E_c$ is an isolated
extended state. Thus, crossing or mergence of curves of $\lambda_{M}(E)/M$
for different $M$'s is a feature for extended states. Equation
(~\ref{finitesizescaling}) has the following asymptotic limits:
$f(x)\propto 1/x$ for $M \to \infty$ and $f(x) \simeq f(0)$ when
$E\to E_c$.

An alternative approach to study localized and extended states is to compute
the participation ratio (PR) of an eigenstate $\psi_{E}(\vec{x})$
with energy eigenvalue $E$. In a lattice geometry, it is defined as
\cite{Pereira,Thouless,Wegner}
\begin{align}
\begin{split}
\mbox{PR}(E)=\frac{1}{N\sum_{i}|\psi_{E}(i)|^{4}},
\end{split}
\label{PR}
\end{align}
where $N$ is the total number of lattice sites and $\psi_{E}(i)$ is the
amplitude of a normalized wave function at site $i$. The PR is of order of
$1/N$ for a maximally localized state ($|\psi_E(i)|=\delta_{i,i_0}$), and of
order of $1$ for a maximally extended (uniform) state
($|\psi_E(i)|=1/\sqrt{N}$). For an extended state whose wave function is a
fractal \cite{xrw} of dimension $D$, its PR should scale with $N$ as
$N^{-1+D/2}$. Note that a localized state scales as $N^{-1}$; namely its
fractal dimension is $D=0$. Numerically, however, the distinction between
a fractal state and a localized state requires calculations on a large
enough lattice.
Here, we will use PR mainly to consistently check the information obtained
from the transfer matrix calculations. More concretely, a local peak of PR($E$)
at an energy $\bar{E}$  indicates that the state $\psi_{\bar{E}}$ is less
localized than its neighboring states $\psi_{E \ne \bar{E}}$.  Based on the
transfer matrix method, we may deduce that the state $\psi_{\bar{E}}$ is extended.
Practically we shall diagonalize the Hamiltonian (\ref{Hamiltonian}) on an
$M\times M$ square lattice with $M$ up to 100 and compute PR$(E)$ for all energy
eigenstates according to the above definition.

\section{ Results}
\label{III}
In the first part of this section we trace the location of the extended state
energies on the lowest LB for a fixed magnetic field and varying disorder
strength $W$, while in the second part we analyze their location for a fixed
disorder and varying magnetic field, and then draw a ``phase diagram" of the
lowest LB in the $W -1/B$ plane. \\
\ \\
\subsection{Fixed $B$ and increasing $W$}

\begin{figure}
  \begin{center}
  {\includegraphics[width=7.76cm]{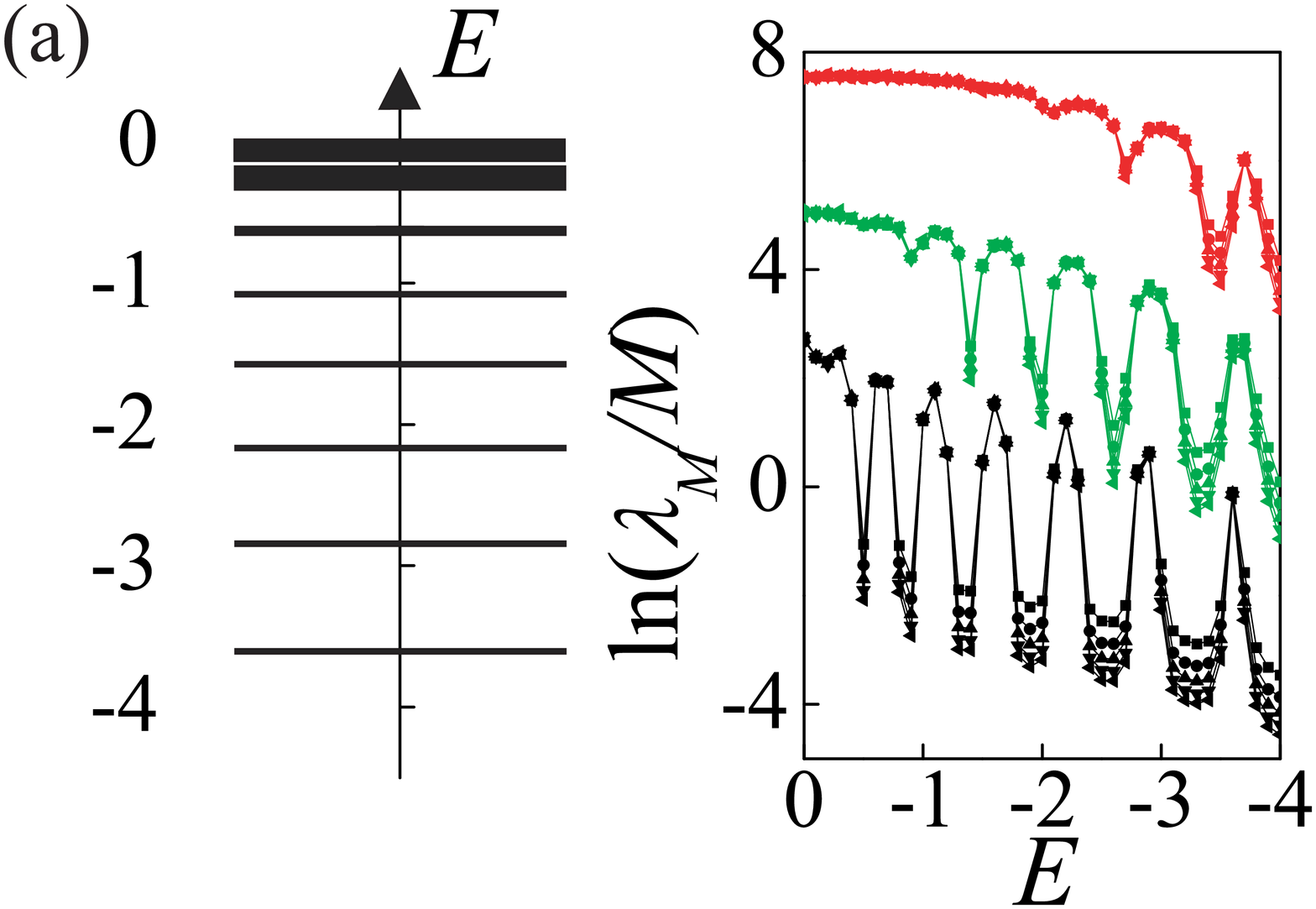}}
  {\includegraphics[width=8cm]{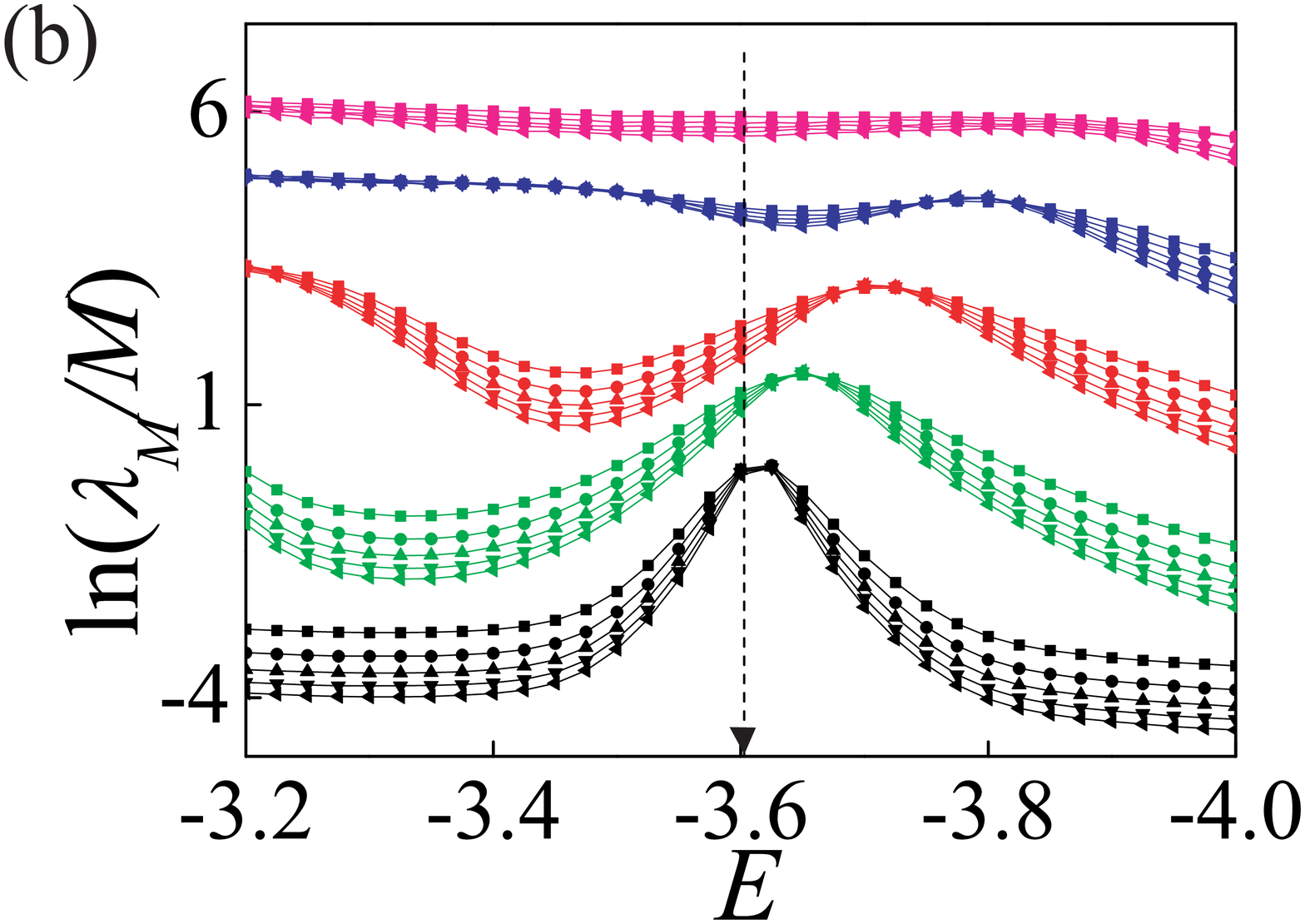}}
  \end{center}
  \caption{(Color online)
  (a) The left panel illustrates the eight Landau subbands in the energy
  range $[-4,0]$ for a clean system for $B=1/15$. The right panel
  displays the quantity $\ln(\lambda_{M}/M)$ as a function of energy $E$ for a
  {\it fixed} magnetic field $B=1/15$ (expressed in quantum flux per square) in
  the same energy range for (from bottom up) $W$=1 (black), 2 (green), 3 (red).
  The numerical data of $\ln(\lambda_{M}/M)$ are obtained by averaging over 40
  samples.
  Bundles of curves for $W=n $ ($n$=2,3) are shifted upward by 2$n$. In each
  bundle, the system widths are $M=32$ (square), 48 (circle), 64 (up-triangle),
  80 (down-triangle) and 96 (left-triangle).
  (b) As in the right panel of (a), but here $\ln(\lambda_{M}/M)$ is displayed in
  the energy range of $[-4,-3.2]$ for (from bottom up) $W$=1 (black), 2 (green),
  3 (red), 4 (blue), 5 (pink). Bundles of curves for $W=n$ ($n$=2,3,4,5) are
  shifted upward by 1.5$n$. The dash arrow indicates the location of extended
  state of a clean system with $B=1/15$.}
\label{AntilevitationWLE}
\end{figure}


First, we focus on the existence and evolution of extended state(s) in the
lowest LB as $B$ (expressed in units of quantum flux per square) is fixed and
$W$ increases. The results of this part elaborate upon earlier ones reported
in Ref.~\onlinecite{Koschny}. For comparison, the left panel of
Fig.~\ref{AntilevitationWLE}(a) illustrates the eight Landau subbands
in the energy range $[-4,0]$ for a clean system with $B=1/15$ since the
electron spectrum is mirror symmetric about $E=0$.
The right panel of Fig.~\ref{AntilevitationWLE}(a) displays the values of
$\ln(\lambda_{M}/M)$ vs $E$ for $B=1/15$, $W=1,2,3,4,5$, and
$M=32,\ 48,\ 64,\ 80,\ 96$ in the same energy range ($E\in[-4,0]$).
Figure~\ref{AntilevitationWLE}(b) displays the  results of
$\ln(\lambda_{M}/M)$ vs. $E$ for $B=1/15$, $W=1,2,3,4,5$, and
$M=32,\ 48,\ 64,\ 80,\ 96$ in a smaller energy range of $E\in[-4,-3.2]$ for
the lowest Landau band (thus achieving higher resolution). Within each bundle
of curves (corresponding to a given value of disorder $W$), the condition
\begin{equation} \label{LocalizedInequality}
M_1>M_2 \ \Rightarrow  \ln(\lambda_{M_1}(E)/M_1)<\ln(\lambda_{M_2}(E)/M_2),
\end{equation}
indicates a localized state at energy $E$. For $W=1,2,3,4$,  the curves in
each bundle for different $M$ merge at the peaks, and at the corresponding
energy $E_c$ the quantity $\ln(\lambda_{M}(E_c)/M)$ is independent of $M$.
At the critical points $E_{c}(W)$, the values of $\lambda_{M}/M$ are nearly
the same besides some numerical errors. Within the one-parameter scaling
ansatz, this indicates quantum Hall transitions between localized states and
isolated extended (critical) states at $E_c$.  The value of $E_c(W)$ depends
on $W$, explicitly $E_{c}(1)=-3.619,\ E_{c}(2)=-3.651,\ E_{c}(3)=-3.718,\ E_{c}(4)=-3.771$.
It should be pointed out that the seemingly merging point near $E=-3.2$ for
$W=3$ and above in Fig.~\ref{AntilevitationWLE}(b) is near the extended
state of the second lowest Landau band.

\begin{figure}
  \begin{center}
  \includegraphics[width=8cm]{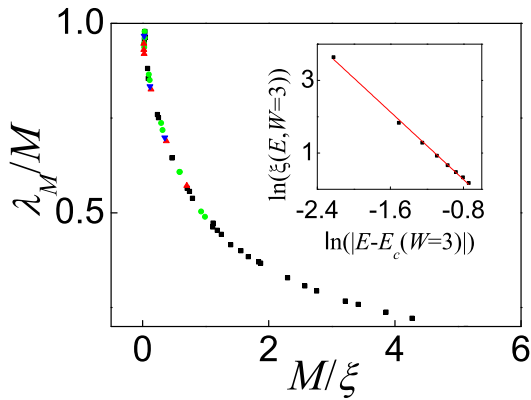}
  \end{center}
\caption{(Color online) The scaling function $\lambda_{M}/M=f(x=M/\xi)$
for $B=1/15$ and $W=1$ (black square), 2 (green circle), 3 (red up-triangle),
4 (blue down-triangle).
The data points are from the critical regime (around the peaks) of
Fig.~\ref{AntilevitationWLE}(b). Inset: The localization length
$\ln(\xi(E,W=3))$ as a function of $\ln(|E-E_{C}(W=3)|)$ for
$E_{c}(W=3)=-3.718$. The solid line is linear fit with slope
$\nu$=2.34.}
\label{AntilevitationWSC}
\end{figure}

To substantiate that quantum phase transitions happen indeed at $E_c$'s,
we show in Fig.~\ref{AntilevitationWSC} that all curves around the peaks of
$W=1,\ 2,\ 3,\ 4$ in Fig.~\ref{AntilevitationWLE}(b) collapse on a single
smooth curve $f(x)$ with $x=M/\xi(E,W)$ when a proper $\xi(E,W)$ is chosen.
Furthermore, $\xi(E,W)$ diverges at $E=E_c(W)$ as  a power law
$\xi \sim [E-E_c(W)]^{-\nu}$. The inset of Fig.~\ref{AntilevitationWSC}
is the curve of $\ln \xi$ vs $\ln (E-E_c)$ for $W=3$ with $E_c=-3.718$.
The nice linear fit with slope $\nu=2.34$ is a strong support of the
one-parameter scaling theory. This value is slightly smaller than the
latest estimate $\nu=2.59$,\cite{Ohtsuki1} but it agrees with the
estimation $\nu=2.34\pm0.04$ \cite{Huckestein2} for the lowest LB.
Thus, from the one-parameter finite-size scaling analysis it is concluded
that the state(s) at $E_c$ for $W=1,\ 2,\ 3,\ 4$ are extended while
states near (but away from) $E_c$ are localized.

While for $W=4$ the peak at $E_c$ at which the curves merge is still
visible (although it is very shallow), we see that for the bundle of curves
corresponding to  $W=5,\ B=1/15$ there is no peak  and the curves
$\lambda_{M}/M$ do not merge. If this bundle of curves for $W=5$ is inspected
at higher resolution, it is found that the inequality
(\ref{LocalizedInequality}) is valid at all energies. This indicates the
{\it absence of Hall transition} for $W=5$. Inspecting $E_c(W)$ from
Fig.~\ref{AntilevitationWLE}(b), we see that, for a fixed magnetic field ($B=1/15$
in this case) the energy of the extended states on the lowest LB
{\it plunges down} as $W$ increases (from $E_{c}=-3.619$ at $W=1$
to $E_{c}=-3.771$ at $W=4$) and then {\it disappears} for a strong enough
disorder ($W=5$). We refer to this slightly downward trend of
$E_c(W)$ on the lowest LB as {\it disorder-driven anti-levitation}.
It contrasts the levitation picture conjectured for continuous systems.
\cite{Laughlin,Khmelnitskii}
It is also slightly distinct from the picture conjectured in previous
works within the lattice geometry,\cite{QNiu1} where it is argued that
$E_c(B,W)=\varepsilon_0(B)$ before the states become localized at higher $W$.

The disappearance of the level $E_c(W)$ of extended states on
the lowest LB at strong disorder raises the question of what
happens with the Chern number attached to that level.
The answer to this question is conjectured in Ref.~\onlinecite{QNiu1}:
At strong disorder $W$ two levels with opposite Chern numbers
approach each other and eventually annihilate each other.
Quantitative substantiation of this conjecture falls beyond the
scope of the present study.

\begin{figure}
  \begin{center}
  \includegraphics[width=8cm]{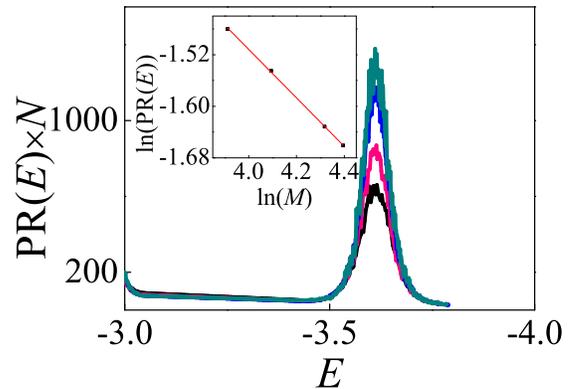}
  \end{center}
\caption{(color online)
Averaged PR$\times N$ as a function of energy $E$ for fixed magnetic
field $B=1/15$ and disorders $W=1$. The lattice size (from down top) is
$51\times 51$ (black), $61\times 61$ (pink), $75\times 75$
(blue), $81\times 81$ (cyan).
The calculation is averaged for 40 samples. Inset: The $\log$(PR) as a
function of $\log (M)$ for $E=-3.621$. The corresponding solid lines are the
linear fit of the data with a slope $-0.39$.
}
\label{AntilevitationWPR}
\end{figure}
Let us now inspect the disorder-driven anti-levitation using the PR method,
that implies the calculation of the wave functions $\psi_{E}(i)$ that
live on an $M\times M$ square lattice, and calculating the relevant PR.
Figure~\ref{AntilevitationWPR} shows PR$\times N$ ($N=M\times M$) as a
function of energy for fixed magnetic field $B=1/15$ and disorder $W=1$
of lattice size of (top down) $51\times 51$ (black), $61\times 61$
(pink), $75 \times 75$ (blue), $81\times 81$ (cyan). Focusing on the first
LB, the highest PR$\times N$ appears at energy $-3.621$, independent of the
sample size, for $W=1$ and $B=1/15$. Unlike the $\lambda_M/M$ curves, the
PR$\times N$ curves for different sample sizes do not cross. The PR$\times N$
peak energies virtually coincide with $E_c(1)=-3.619$ obtained within the
transfer matrix method. Thus, the energy of the highest PR$\times N$ in the
first LB is consistent with $E_{c}(W)$ discussed within the one-parameter
finite size scaling hypothesis. One can further see that PR$\times N$ peaks
indeed correspond to extended states by studying the sample size dependence
of peak heights and PR$\times N$ at other energies. Focusing on
Fig.~\ref{AntilevitationWPR}, first one can clearly see that at the energy far
from the peak energy $E_{c}(1)=-3.621$ the PR$\times N$ is $M$ independent
which implies $D=0$ for these energies. Namely, these states are localized.
Second, at the peak energy $E_{c}(1)=-3.621$, the PR$\times N$ increases
with $M$. The inset of Fig.~\ref{AntilevitationWPR} is the natural logarithm
plot of the PR vs sample size $M$ at the peak with an exponent of $-0.39$,
indicating a fractal wave function of dimension $D=2-0.39=1.61$ for the peak.
This result is consistent with the multifractal analysis of the integer
quantum Hall effect \cite{Huckestein3} where it is found that the fractal
dimension of extended states on the first LB is $D=1.6$.
Third, at energies slightly different from the peak $E_{c}(1)=-3.621$
the curves PR$\times N$ do not merge for small $M$ ($M=51$ and $M=61$)
but merge for larger $M$ ($M=75$ and $M=81$). For a large enough system the
PR$\times N$ should merge together for different $M$ at all energy
(localized states) except for the peak (extended states).
Since the energy of the peak is size-independent, anti-levitation can be derived
without resorting to finite size scaling analysis. However, for the calculation
of the critical exponent $\nu$ one must employ finite-size scaling analysis either
within transfer matrix formalism or within PR analysis.
\\

In Ref.~\onlinecite{Koschny} the authors calculate the Hall conductivity
and the localization length for white noise and also for short range
correlated disorder. For the white-noise disorder the disorder-driven
anti-levitation scenario is found while for the finite-range correlated
disorder a weak levitation is noticed. An indirect substantiation of the
disorder-driven anti-levitation scenario for Gaussian white noise on-site
potential can also be found by analyzing the results in 
Ref.~\onlinecite{Prodan}. The authors calculated $\sigma_{xx}$ and 
$\sigma_{xy}$ for a lattice geometry. By inspecting their results at $B=1/6$ 
it can be seen that the first plateau transition occurs at an energy 
$E_c \approx -3.18$ that is lower than the energy of the lowest Landau level 
in a clean system ($\varepsilon_0 \approx -3.09$).

\begin{figure}
  \begin{center}
  \includegraphics[width=6.512cm]{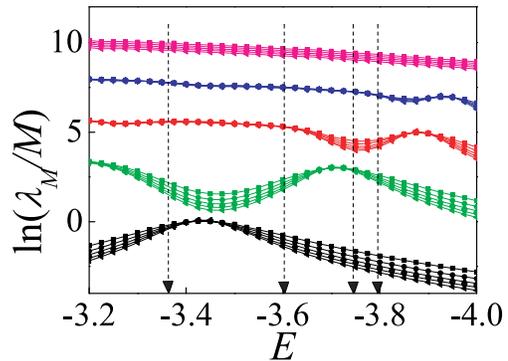}
  \end{center}
\caption{(Color online) The figure displays $\ln(\lambda_{M}/M)$
  (averaged over $40$ samples) as a function of energy $E$ for fixed disorder
  strength $W=3$ at different magnetic fields. The curve bundles from bottom
  up correspond to $B=1/9$ (black), $1/15$ (green), $1/24$ (red), $1/30$ (blue),
  and $1/40$ (pink). In order to have a better view, bundle of curves of
  $B=1/15$ is shifted upward by 3 relative to those of $B=1/9$. The bundles of
  $B=1/24$, $B=1/30$, and $B=1/40$ are then shifted upward by 2 in order. In each
  bundle, the system width is $M=32$ (square), $48$ (circle), $64$
  (upper triangle), $80$ (down triangle), $96$ (left triangle).
  The dashed arrows indicate the locations of extended states (from left to right)
  of clean systems with $B=1/9$, 1/15, 1/24, and 1/30, respectively.}
\label{AntilevitationBLE}
\end{figure}
\ \\
\subsection{Fixed $W$ and decreasing $B$}

\begin{figure}
  \begin{center}
  \includegraphics[width=8cm]{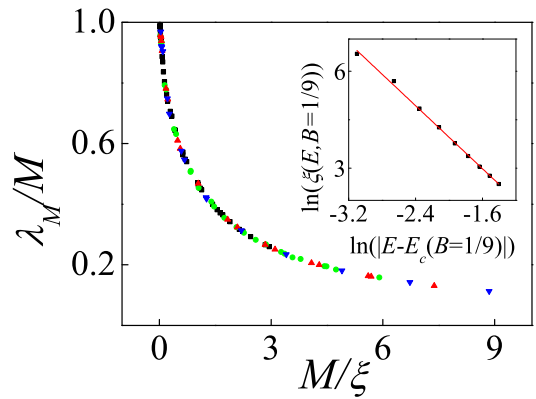}
  \end{center}
\caption{(Color online) The scaling function $\lambda_{M}/M=f(x=M/\xi)$
  for $W=3$ and $B=1/9$ (black square) ,1/15 (green circle) ,1/24
  (red up-triangle) ,1/30 (blue down-triangle).
The data points are from the critical regime (around the peaks) of
Fig.~\ref{AntilevitationBLE}. Inset: The localization length
$\ln(\xi(E,B=1/9)$ as a function of $\ln(|E-E_{C}(B=1/9)|)$ for
$E_{C}(B=1/9)=-3.444$.
The solid line is linear fit with slope $\nu$=2.34.
  }
\label{AntilevitationBSC}
\end{figure}

\begin{figure}
  \begin{center}
  \includegraphics[width=6.512cm]{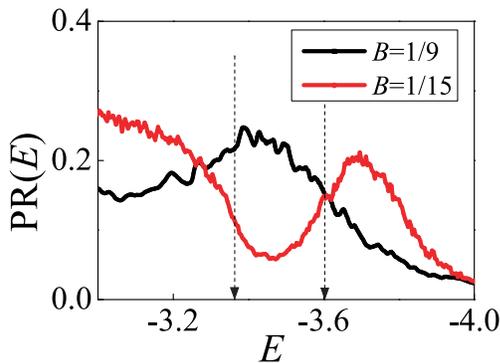}
  \end{center}
\caption{(Color online)
  Averaged PR as function of energy $E$ for a fixed disorder
  ($W=3$) and two different green magnetic fields $B=1/9$ (square)
  and $B=1/15$ (circle).
  The lattice size is chosen as $81\times81$ square lattice.
  The calculation is averaged for 40 samples.
  The dash arrows indicate the locations of extended states of a clean
  system with $B=1/9$ (left) and $B=1/15$ (right).}
\label{AntilevitationBPR}
\end{figure}

Next, we address the effect of decreasing magnetic field on the first
extended state(s) at a fixed moderate disorder. Assuming the effect of
increasing disorder and lowering magnetic field on the behavior of energies
of extended states enters through the dimensionless parameter $\omega_c \tau$
it is natural to inspect anti-levitation at fixed $W$ and decreasing $B$.
However, we are unaware of similar analysis, (for example, in
Ref.~\onlinecite{Koschny} $B$ is fixed and $W$ is changed).
In Fig.~\ref{AntilevitationBLE}, the average $\ln(\lambda_{M}/M)$ is displayed
versus energies for $W=3$, different system widths $M=32,\ 48,\ 64,\ 80,\ 96,$
and magnetic fields $B=1/9,\ 1/15,\ 1/24,\ 1/30$. Curve bundles for $B>1/30$
display peaks at which $\ln(\lambda_{M}/M)$ are merged for different $M$.
At the merge points, the values of $\lambda_{M}/M$ of different bundles are the
same. The mergence is confirmed by the finite-size scaling analysis
that all data around the peaks collapse onto a single smooth curve,
$\lambda_{M}/M=f(x=M/\xi(E,B))$, as shown in Fig.~\ref{AntilevitationBSC}
when a proper $\xi(E,B)$ is used. Indeed the scaling functions in
Fig.~\ref{AntilevitationWSC} and Fig.~\ref{AntilevitationBSC} are exactly the
same (they overlap with each other) and they are also the same as in Fig.~15 of
Ref.~\onlinecite{Huckestein}, implying the scaling function is universal in the
integer quantum Hall system. Furthermore, the extracted $\xi(E,B)$ diverges in
a power-law fashion at energy $E_c$ whose value changes with $B$ as shown in
the inset of Fig.~\ref{AntilevitationBSC} with the critical exponent of $\nu=2.34$,
the same as the one found earlier. The numerical critical energies are $E_{c}=-3.444$
for $B=1/9$, $-3.712$ for $B=1/15$, $-3.901$ for $B=1/24$, and $-3.955$ for $B=1/30$.
The corresponding states at these energies are extended (or, more precisely, critical).
The values of $E_{c}=-3.404$ for $B=1/9$ and $E_c=-3.718$ for $B=1/15$
are consistent with those at the peak positions obtained through
the PR calculations shown in Fig.~\ref{AntilevitationBPR}.
\ \\

\begin{figure}
  \begin{center} 
  \includegraphics[width=8cm]{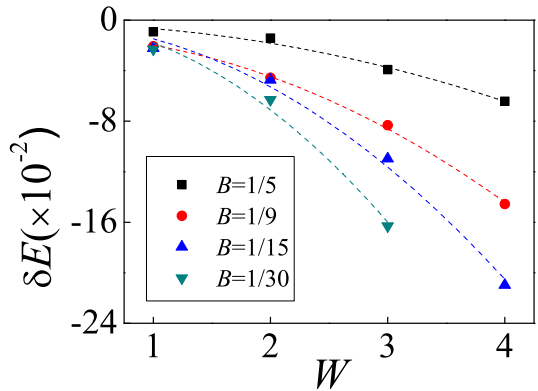}
  \end{center}
  \caption{(Color online) Energy deviations between the extended level and
  the center of the first LB $\delta E \equiv E_c(B,W)-\varepsilon_0(B)$
  are shown as function of disorder $W$. The curves from top to bottom
  correspond to $B=1/5,\ 1/9, \ 1/15$, and $1/30$. Dashed
  lines are the fits of quadratic functions $-a(B)W^2$.
  $a(B)$ increases as $B$ decreases.
  }
\label{dEW}
\end{figure}

Remarkably, at lower magnetic field, for example $B=1/40$ in
Fig.~\ref{AntilevitationBLE}, the corresponding bundle of curves does
not have a peak, and when inspected with higher resolution, its curves
for different $M$ follow inequality (\ref{LocalizedInequality}).
In other words, there is a critical (disorder dependent) magnetic field
$B_c(W)$ below which the extended states on the lowest LB become localized.
This is qualitatively consistent with the results of Ref.~\onlinecite{QNiu1}.
There is, however a difference between our results and those of
Ref.~\onlinecite{QNiu1} regarding the behavior of the critical energy
$E_c(B,W)$ for $B>B_c(W)$. We arrive at the somewhat unexpected result that
with decreasing magnetic field $E_{c} (B,W)$  plunges down faster than
$\varepsilon_0(B)$ (the lowest Landau level in Hofstadter's butterfly
\cite{Hofstadter} that is a decreasing function of $B$). In short, for
$B \ge B_c(W)$ we find $E_{c} (B,W)<\varepsilon_0(B)$ and eventually,
for $B < B_c(W)$, there are no extended states on the lowest LB.
We refer to this scenario as {\it magnetic-field-driven anti-levitation}.
The qualitative explanation of Chern number annihilation mechanism
suggested in Ref.~\onlinecite{QNiu1} applies here as well.
The essential results of our extensive numerical calculations are
displayed in Fig.~\ref{dEW}, which shows the deviation
$E_c(B,W)-\varepsilon_0(B)$ between the first extended state(s) energy
and the center of the first LB with varying disorder and magnetic field.
An obvious anti-levitation of the first extended state(s) energy $E_c(B,W)$
can be observed at strong disorder and/or for weak magnetic field. \\

The magnetic-field-driven anti-levitation can be understood following
the principle of level repulsion or avoided crossing. Assume $E_c$ is
the energy of the extended state in the lowest LB for a fixed field $B$;
when the random potential of zero mean increases by $\Delta V$, the
energy shift of the extended state at the second-order perturbation is
\begin{align}
\begin{split}
\Delta E_{c} =\sum_{E\ne E_c}
\frac{|\langle E|\Delta V|E_c\rangle|^{2}}{E_c-E},
\end{split}
\label{Landaulevelenergy}
\end{align}
Here $|E_c\rangle$ is an extended state on the lowest LB at energy
$E_c$ and $|E\rangle$ denotes an arbitrary state (possibly localized,
and including higher Landau bands) at energy $E\neq E_c$. Note that both 
$|E_c\rangle$ and $|E\rangle$ correspond to the ``unperturbed'' system at 
potential $V$. We also assume that $V$ is strong enough to lift the 
degeneracy of the lowest LB, justifying the use of non degenerate 
perturbation theory. It is expected that the contribution from localized 
states will be much smaller than that of the extended ones, and
therefore, assuming that the states $E_c$ are extended and
belong to higher LB. Since $E_c$ is located around the LB
center, there are more states whose energies are above
$E_c$ than those below $E_c$. Thus, more terms in the sum
are negative, and $\Delta E_c<0$, implying anti-levitation. According
to Eq. \eqref{Landaulevelenergy}, the shift should be proportional to
$W^2$. This is indeed consistent with our numerical data
points that agree with quadratic fits $\Delta E_c=-a(B)W^2$
(dashed lines in Fig.~\ref{dEW}). As for the dependence of $\Delta E$ on
$B$ for fixed $W$, we note that the denominator on the right-hand side of 
Eq. \eqref{Landaulevelenergy} is approximately proportional to the
Landau level spacing (recall that $|E\rangle$ belongs to higher
LB). This suggests an estimate $a(B)\sim1/B$. In the inset
of Fig.~\ref{dEW} we plot $a(B)$ vs $1/B$. The deviation from
straight line is apparently due to the dependence of the
matrix element $\langle E_c|\Delta V|E\rangle$ on $B$, which is 
difficult to elucidate (remember that the wave functions correspond
to systems with strong disorder), but appears to be small
especially for small magnetic fields.

\begin{figure}
  \begin{center} 
  \includegraphics[width=8cm]{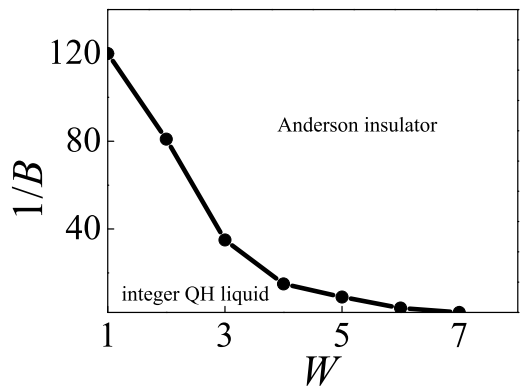}
  \end{center}
\caption{Phase boundary between the integer
QH liquid and the Anderson insulator on $W-1/B$ plane.
}
\label{PB}
\end{figure}


Similar anti-levitation is also observed for the extended states in the
second lowest LB, but it is less pronounced than that of the lowest LB
as commensurate with the principle of level repulsion.
The disappearance of the lowest extended state at very small magnetic
field indicates a transition between the IQHE state on the lowest LB and
an Anderson insulator.

Based on our analysis pertaining to the lowest LB, Fig.~\ref{PB}
displays a phase diagram in the $W-1/B$ plane where the boundary between
the integer quantum Hall liquid and Anderson insulator is marked.
In principle, the line separating the two phases approaches infinity on
the $1/B$ axis as $W\to 0$. It seems to end at $W \approx 7$ on the $W$
axis, implying that no extended state exists beyond this level of disorder.
Based on a pertinent experiment,\cite{Huang}, a similar phase diagram is
established albeit for higher Landau bands $2<n<10$. Our phase diagram is
consistent with the experimental one and extends it to the lowest LB.

\section{Conclusions}
\label{IV}
In this work we numerically studied the behavior of extended
state energies $E_c(B,W)$ in the lowest LB for an electron on 
a square lattice with white-noise disorder of
strength $W$ subject to an external (perpendicular) magnetic
field of strength $B$. It is found that $E_c(B,W)$ exhibits
disorder- and magnetic-field-driven anti-levitation
typically for $W > \hbar \omega_c$. Concretely, for fixed magnetic
field, $E_c(B,W)$ plunges down with the estimate
$E_c(B,W)-E_c(B,0)\sim -a(B)W^2$, with $a(B)\sim1/B$ (at
least for weak magnetic field). This scenario may not
exist for long-range correlated disorder \cite{Koschny}. The fact
that for a fixed disorder strength $W$, $E_c(B,W)$ plunges
down and eventually disappears as the magnetic field decreases
is explained in Ref. \onlinecite{QNiu1}, based on annihilation of
two levels carrying Chern number with opposite signs. A
phase diagram of the IQHE is drawn in the $W-1/B$ plane
where a clear boundary is identified which distinguishes
the IQH liquid and an Anderson insulator.

\section{Acknowledgement}

C. W. thanks T. Ohtsuki for many helpful suggestions concerning the
numerical calculations and M. Ma for valuable discussions.
Y. A. thanks E. Prodan for discussing the interpretation of his results
in Ref.~\onlinecite{Prodan}. This work is supported by Hong Kong
RGC Grants (No. 604109 and No. 605413) and NNSF of China Grant (No. 11374249).
The research of Y. A. is partially supported by Israeli Science
Foundation Grants No. 1173/2008 and No. 400/2012.

\end{document}